\documentstyle[12pt]{article}
\begin{document}
\baselineskip = 6.0mm
\topmargin= -5mm
\begin{center}
\begin{Large}

 No Goldstone boson in  NJL  and Thirring models  

\end{Large}

\vspace{1cm}

T. Fujita and M. Hiramoto 

Department of Physics, Faculty of Science and Technology, 
\\ Nihon University, Tokyo, Japan \\

and  

H. Takahashi

Laboratory of Physics, Faculty of Science and Technology, \\ 
Nihon University, Chiba, Japan

\vspace{1cm}

{\large Abstract}

\end{center}

\vspace{1cm}

We present an intuitive picture of the 
chiral symmetry breaking in the NJL and Thirring  models. 
For the current current interaction model with the massless fermion, 
the vacuum is realized with the chiral symmetry broken phase, 
and the fermion acquires the induced mass.  With this finite 
fermion mass, we calculate the boson mass, and find that 
the boson is always massive, and there exists no Goldstone boson  
in the Nambu-Jona-Lasinio (NJL) model.

\newpage

\section{ Introduction}

In field theory, physical vacuum 
can violate the symmetry which is possessed in the Lagrangian. 
In this spontaneous symmetry breaking, there is the Goldstone theorem   \cite{q1,q2} 
which states that there appears a massless boson if the symmetry is 
spontaneously broken. In the boson field theory models, 
the existence of a massless boson may well be justified since bosons are found 
from the beginning in the models. 

On the other hand, fermion field theory models are quite different in that 
bosons must be constructed by fermions and antifermions by their dynamics.  For the 
proof of the existence of the massless boson, therefore, one first has to find 
the degrees of freedom that correspond to the boson, and this degrees 
of freedom should be obtained from the fermion and antifermion degrees 
of freedom. This is indeed a nontrivial task, and these models 
which we discuss below such as the NJL and Thirring  models cannot 
find any degrees of freedom for the massless boson 
after the spontaneous symmetry breaking. Therefore, there exists 
no Goldstone boson for the NJL and Thirring  models. 

In this paper, we present a clear and intuitive picture why  
there appears no Goldstone boson and how the symmetry is broken in the vacuum 
in these fermion field theory models. In particular, we  show 
an essential difference between boson and fermion field theory. 
For the NJL and the Thirring models, we describe some basic physical quantities 
in terms of a  few simplified  equations.  In the chiral symmetry breaking, 
the fermion naturally acquires the mass, and the vacuum structure is determined, 
and it has a finite condensate. 
But there is no freedom left for the massless boson, and the rest of the theory 
becomes just the massive fermion field theory with the same number of the fermion 
degrees of freedom as before the symmetry breaking. Therefore, 
the problem of the chiral symmetry breaking 
is at this point completed.  Whether the rest 
of the massive fermion field theory may have an exponentially small mass 
or a heavy massive boson 
is irrelevant to the symmetry breaking business. In reality, 
the NJL and the Thirring models have one massive boson since 
they are essentially a delta function potential in coordinate space.

In addition, we should note that most of the calculated results here 
on the NJL model are just the same as those of their 
original paper \cite{q7}. The only but essential difference 
between their calculations and ours is concerned 
with the mass of the boson. They evaluated the boson mass without 
solving the Bethe-Salpeter equations, but speculated it with the pole of the S-matrix 
which does not necessarily correspond to a bound state. On the other hand, 
we calculate the mass of the boson, solving the dynamics properly. 

This paper is organized as follows. 
In the next section, we briefly review the spontaneous symmetry breaking  
in boson field theory. In section 3, we present  an intuitive picture 
of the symmetry breaking in fermion field theory. Section 4 treats 
the symmetry breaking in terms of the Bogoliubov transformation. 
In section 5, we discuss how to evaluate the boson mass, and 
in section 6, we summarize what we clarify in this paper.

\vspace{0.5cm}

\section{ Spontaneous symmetry breaking in boson field theory }

Here, we discuss the spontaneous symmetry breaking in boson 
field theory. This can be found in any field theory text books, and therefore 
we only sketch the simple picture why the massless boson appears 
in the spontaneous symmetry breaking. 

The Hamiltonian density for complex boson fields can be written as 
$$ {\cal H} = {1\over 2} ({\bf p}{\phi}^{\dagger})({\bf p}{\phi})+
U\left( |{\phi}| \right) . 
\eqno{(2.1)} $$
This has a $U(1)$ symmetry, and when one takes the potential as
$$ U\left( |{\phi}| \right) = U_0 \left(  |{\phi}|^2 
-\lambda^2 \right)^2 \eqno{(2.2)} $$
then, the minimum of the potential $ U\left(  |{\phi}| \right)$ 
can be found at $|\phi|=\lambda$. But one must notice that this is 
a minimum of the potential, but not the minimum of the total energy. 

The minimum of the total energy must be found together with the kinetic 
energy term. 
When one rewrites the complex field as 
$$ \phi = (\lambda + \rho) e^{i{\xi\over{\lambda}}} \eqno{(2.3)} $$
then, one can rewrite eq.(2.1) as
$$ {\cal H} = {1\over 2}\left[ ({\bf p}{\xi})({\bf p}{\xi})
+({\bf p}{\rho})({\bf p}{\rho})
\right]+U\left( |\lambda + \rho| \right) +... \eqno{(2.4)} $$
Here, one finds the massless boson $\xi$ which is associated with the degeneracy  
of the vacuum energy. 
The important point is that this infinite degeneracy of the potential vacuum 
is converted into the massless boson degrees of freedom when 
the degeneracy of the potential vacuum is resolved by the kinetic energy term. 

\vspace{0.5cm}

\section{ Chiral symmetry breaking in fermion field theory }

Now, we present an intuitive discussion of the chiral symmetry 
breaking in the fermion field theory models. 
Here, our discussions are restricted to the current-current interaction, and 
thus we treat the  NJL  model. The Thirring model can be treated just 
in the same way. 

The Hamiltonian density of the NJL model is written as 
$$ {\cal H}= \psi^{\dagger} {\bf p}\cdot{\mbox{\boldmath $\alpha$}}
\psi -{1\over 2}G \left[ (\bar{\psi}\psi )^2 
+(\bar{\psi}i\gamma_5\psi )^2  \right] . \eqno{(3.1)}  $$
Here, we take the chiral representation, and denote the $\psi$ as 
$$ \psi({\bf n},s)=\frac{1}{\sqrt{2}}
\left( \begin{array}{c}
 \psi_1{{\mbox{\boldmath $\sigma$}} \cdot \bf{\hat n}}\chi^{(s)} \\
      \psi_2 \chi^{(s)} \end{array}
\right) \eqno{(3.2)} $$
where  $\chi^{(s)}$ denotes the spin part. 
In this case, the Hamiltonian density for the NJL model 
can be written afer the summation of $s$ is taken 
$$ {\cal H}= \psi_1^{\dagger} ({\bf p}\cdot{\bf{\hat n}}) \psi_1 
-\psi_2^{\dagger} ({\bf p}\cdot{\bf{\hat n}}) \psi_2
 + 2G ( \psi_1^{\dagger} \psi_1 \psi_2^{\dagger} \psi_2 ) .  \eqno{(3.3)}  $$
In the same way as the boson case, we can define the potential $U(\psi_1,\psi_2)$ as 
$$ U(\psi_1,\psi_2) = 2G|\psi_1|^2|\psi_2|^2 . \eqno{(3.4)} $$
It is clear from this equation that the potential of the fermion 
field theory models does not have any nontrivial minimum, apart from the trivial one 
$ \psi_1^{\dagger} \psi_1=0$, $\psi_2^{\dagger} \psi_2=0$. 
This is in contrast to the boson case where there is a 
nontrivial minimum in the potential. 
Therefore, there is no degeneracy of the true vacuum state since the minimum 
of the potential here is only a trivial one. 

Where can one find the new vacuum that violates the chiral 
symmetry ? The answer is simple. One has to consider the kinetic 
energy term. In the fermion system, the kinetic energy is negative 
for the vacuum state. 
In what follows, we present a simple and intuitive argument of obtaining a new 
vacuum state including the kinetic energy term. This treatment is schematic, 
but one can learn the essence of the physics of the chiral symmetry 
breaking in fermion field theory. The exact treatment 
will be given in section 4. 

Now, we can take an average value  
of the kinetic energy ($-\Lambda_0$) for the negative energy state, 
and thus we write eq.(3.3) as
$$ {\cal H} \approx -\Lambda_0 \left(|\psi_1|^2 +|\psi_2|^2 \right)
 + 2G  |\psi_1|^2 | \psi_2 |^2  .  \eqno{(3.5)}  $$
This can be rewritten as 
$$ {\cal H}=  2G \left( |\psi_1|^2-{\Lambda_0\over{2G}} \right) 
\left( |\psi_2|^2-{\Lambda_0\over{2G}} \right)
-{\Lambda_0^2\over{2G}} .   \eqno{(3.6)}  $$
Therefore, it is easy to find the $|\psi_1|^2$ and 
$|\psi_2|^2 $ for the new vacuum state, that is, 
$$  |\psi_1|^2=  {\Lambda_0\over{2G}},  \qquad
  |\psi_2|^2=  {\Lambda_0\over{2G}}  . \eqno{(3.7)} $$
In this case, the vacuum energy $E_{vac}$ 
and the condensate $C$ become 
$$ E_{vac}= -{\Lambda_0^2V\over{2G}} \eqno{(3.8a)}   $$
$$ C={\Lambda_0\over G}  \eqno{(3.8b)} $$
where $V$ denotes the volume of the system. 
Now, we expand the $\psi_1$ and $\psi_2$ in terms of the minimum state [eqs.(3.7)] as
$$  \psi_1=\sqrt{{\Lambda_0\over{2G}}}{\eta} + \tilde{\psi_1}, \qquad 
  \psi_2=\sqrt{{\Lambda_0\over{2G}}}{\eta} + \tilde{\psi_2}  \eqno{(3.9)} $$
where $\eta$ and $\bar{\eta}$ denote grassmann numbers which satisfy 
the following algebras
$$ \eta \bar{\eta}=1,  
\qquad \eta \eta =0 , \qquad \eta \tilde{\psi_i} = -\tilde{\psi_i} \eta . \eqno{(3.10)} 
$$
In this case, the Hamiltonian density becomes
$$ {\cal H}= -{\Lambda_0^2\over{2G}} + 
\Lambda_0\left( \tilde{\psi}_1^\dagger\tilde{\psi_2} 
+h.c. \right) $$
$$+ \tilde{ \psi}_1^{\dagger} ({\bf p}\cdot{\bf{\hat n}}) \tilde{ \psi}_1
 -\tilde{\psi}_2^{\dagger} ({\bf p}\cdot{\bf{\hat n}}) 
 \tilde{\psi}_2+ 2G  |\tilde{\psi}_1|^2 |\tilde{ \psi}_2 |^2+ 
O(\tilde{ \psi}_1, \tilde{\psi}_2^{\dagger}) .   \eqno{(3.11)}  $$
Now, it is clear that the second term is the mass term. Therefore, one notices 
that after the chiral symmetry breaking, the fermion acquires the finite mass, 
and the induced 
mass $M$ becomes $M=\Lambda_0$. Therefore, at this point, the symmetry breaking 
problem is completed. The rest of the field theory becomes just the massive 
fermion field theory. For example, the Thirring model becomes the massive 
Thirring model where one knows well that there exists one massive boson, and 
the mass spectrum is obtained as the function of the coupling constant \cite
{q11,q12,q13,q14}. 

This means that one cannot find a massless 
boson in the Hamiltonian of the fermion system. It is also quite important 
to note that the new Hamiltonian is still described by the same number 
of the fermion degrees of freedom as the original one. 
This is in contrast to the boson case 
where one of the complex field freedom becomes the massless boson $\xi$. 

Therefore, if one wants to find any boson in the NJL model, 
then one has to solve the dynamics since the kinematics cannot produce any Goldstone 
boson 
in fermion field theory. However, intuitively, 
one sees that it should be difficult to find 
a massless boson as a bound state of fermion and antifermion system, 
regardless the strength of the coupling constant in the system of the finite fermion 
mass.   
Indeed$B!$(Bthere is no massless boson in the NJL as well as the massless Thirring models 
if one solves the dynamics properly as will be seen below in the next section. 

Here,  it is interesting to note that the vacuum energy 
and the condensate with this value of the $\Lambda_0$ [eq.(3.8$a$) and eq.(3.8$b$)] 
become 
$$ E_{vac}= -{M^2V\over{2G}}  \eqno{(3.12a)} $$  
$$ C={M\over G} \eqno{(3.12b)} $$ 
which are quite close to the exact calculations. 
Also, the chiral charge $Q_5$ of the vacuum can be evaluated and is found to be 
$$ Q_5 =0  \eqno{(3.13)} $$
which is also consistent with the exact one.

We summarize the intuitive discussions here for the fermion field 
theory. Even though there is no nontrivial minimum in the potential, 
one finds a new vacuum if one considers the kinetic energy terms. 
The chiral symmetry is broken in the new vacuum state 
of the NJL and the Thirring  models. 
After the symmetry breaking, the fermion acquires the mass. 
But there is no massless boson since the degree of freedom 
for the massless boson does not exist. The mass of the boson predicted in the field 
theory of the finite fermion mass has therefore nothing to do with the symmetry breaking 
business.

\vspace{0.5cm}

\section{ Bogoliubov transformation of  NJL model }

In this section, we discuss the chiral symmetry breaking 
in the Nambu-Jona-Lasinio model in terms of  
the Bogoliubov transformation of the Hamiltonian. 
Since this treatment is already given in detail for the NJL model 
by Nambu and Jona-Lasinio \cite{q7}, we discuss only 
briefly the main results of the above fermion field theory model 
of current current interaction.  

Now, we define new fermion operators by the Bogoliubov transformation, 
$$ c({\bf n},s) = 
\cos \left({\theta_{\bf n}\over 2}-{\pi\over 4}\right) a({\bf n},s) 
- s \sin \left({\theta_{\bf n}\over 2}-{\pi\over 4}\right) 
b^{\dagger}(-{\bf n},s) \eqno{(4.1a)} $$
$$ d^{\dagger}(-{\bf n},s) =  
\cos \left({\theta_{\bf n}\over 2}-{\pi\over 4}\right) b^{\dagger}(-{\bf n},s) + 
s \sin \left({\theta_{\bf n}\over 2}-{\pi\over 4}\right) a({\bf n},s) \eqno{(4.1b)} $$
and, we can obtain the new Hamiltonian under the Bogoliubov transformation, 
$$ H= \sum_{{\bf n},s}\left\{ |{\bf p}_{\bf n}| \sin{\theta_{\bf n}} 
+\frac{2G}{L^3}{\mathcal{B}}
\cos{\theta_{\bf n}}\right\}   
 \left( c^{\dagger}({\bf n},s)c({\bf n},s)+d^{\dagger}(-{\bf n},s)d(-{\bf n},s) 
\right) $$
$$ + \sum_{{\bf n},s}
\left\{-|{\bf p}_{\bf n}|s \cos{\theta_{\bf n}}+
\frac{2G}{L^3}{\mathcal{B}} s \sin{\theta_{\bf n}} \right\}  
\left(c^{\dagger}({\bf n},s)d^{\dagger}(-{\bf n},s)+
d(-{\bf n},s)c({\bf n},s) \right)  + {H'}_{int}  \eqno{ (4.2)} $$
where $ {H'}_{int}$ is the interaction term. 
${\mathcal{B}} $ is defined as 
$$ {\mathcal{B}}=\sum_{{\bf n},s}\cos\theta_{\bf n} .\eqno{ (4.3)} $$ 

Now, we define the new fermion mass $M$ by
$$ M = \frac{2G}{L^3}\mathcal{B}. \eqno{ (4.4)} $$
The Bogoliubov angle $\theta_{\bf n}$ can be determined from the following 
equation
$$ \tan \theta_{\bf n}  = {|{\bf p}_{\bf n}| \over M} . \eqno{ (4.5)} $$
In this case, the vacuum changes drastically since the original 
vacuum is trivial. From eqs. (4.3)(4.4)(4.5), one can express the fermion 
mass $M$ in terms of the cut off meomentum $\Lambda$. The vacuum energy and 
the condensate are also calculated \cite{q07}, and are close to 
the ones given in eqs. (3.12). 

\vspace{0.5cm}

\section{ Boson Mass}

Now, we calculate the boson mass in the NJL model. 
For the NJL model, there are many calculations 
of the boson mass \cite{q07}. All of the calculations show that 
the boson mass is always finite when the fermion mass is finite. 

Now, a question is why people found a massless boson. This is rather clearly 
seen. Namely, the fermion mass is set to zero again, and in this case, one 
obtains a mssless boson since the boson mass is measured in units of 
the fermion mass $M$. But as we see above, we cannot set the fermion mass to zero 
after the symmetry breaking since the vacuum with the induced fermion mass  
is lower than the trivial one, and thus is realized in nature. 

Below we show the calculated boson mass with the fermion mass $M$, and 
it is clear that we can only obtain a finite boson mass. 

The boson state $|B\rangle$  can be expressed as
$$ |B\rangle = \sum_{{\bf n},s}f_{\bf n}
c^{\dagger}({\bf n},s)d^{\dagger}(-{\bf n},s)|\Omega\rangle, 
 \eqno{(5.1)} $$
where $f_{\bf n}$ is a wave function in momentum space, and $|\Omega\rangle$ 
denotes the Bogoliubov vacuum state.
The equation for the boson mass $ {\cal M}$ for the NJL model  is written 
in terms of the Fock space expansion at the large $L$ limit 
$$ {\cal M}f(p) = 2E_{p}f(p)
-\frac{2G}{(2\pi)^3}\int^{\Lambda} d^3q  f(q)\left(1+
\frac{M^2}{E_{p}E_{q}}+\frac{{\bf p}\cdot {\bf q}}{E_{p}E_{q}}\right)  . 
 \eqno{(5.2)} $$
This equation can be easily solved since it is a separable type, and 
we can obtain the boson mass as the function of the coupling constant $G$ \cite{q9,q10}. 
The boson mass is always finite, and there is no bound state if the coupling 
constant $G$ is smaller than the critical value. 
 
\vspace{0.5cm}

\section{ Conclusions}

We have presented a clear and intuitive picture of the chiral symmetry 
breaking in the NJL and the Thirring models. In these fermion field theory models, 
we have learned that the chiral symmetry is indeed broken and the new vacuum is 
determined. The vacuum has a chiral condensate and the fermion acquires 
the mass after the symmetry breaking. The symmetry breaking problem is at 
this point completed since the number of the fermion degrees of freedom does not change. 
Even though  the NJL as well as the Thirring models have a massive boson, the 
question of the boson mass is not relevant to the symmetry breaking business. 

Therefore, the concept of the Goldstone boson in these fermion models should be 
discarded. The question of the boson mass spectrum after the symmetry breaking 
in eq. (3.11) for the fermion system is just the same as asking 
what should be the excitation spectrum of the boson 
of $\rho$ in eq. (2.4) in the boson system.

\end{document}